\documentclass[%
 reprint,
superscriptaddress,
  prl
]{revtex4-2}

\usepackage{graphicx}
\usepackage{dcolumn}
\usepackage{bm}
\usepackage{float}
\usepackage{hyperref}
\usepackage[mathlines]{lineno}
\usepackage{lipsum}
\usepackage[english]{babel}
\usepackage{physics}
\usepackage{float}
\usepackage{fancyhdr}
\usepackage{booktabs}
\usepackage{mathrsfs}
\usepackage{dsfont}
\usepackage{mathrsfs,amssymb, amsmath, amsbsy}
\usepackage{xcolor}

\newcommand{\jb}{\overline{\jmath}}
\newcommand{\mb}{\overline{m}}

\begin{document}

\preprint{APS/123-QED}

\title{The quantum beam splitter with many partially indistinguishable photons: multiphotonic interference and asymptotic classical correspondence  }

\author{Miguel E. Villalobos}
\email{me.villalobos@uniandes.edu.co}
 \affiliation{Quantum Optics Group, Physics Department, Universidad de los Andes, Bogot\'{a} 111711, Colombia}
 \affiliation{Physics Department, Politecnico di Milano, Milan 20133, Italy}

\author{Alejandra Valencia}
\email{ac.valencia@uniandes.edu.co}
\affiliation{Quantum Optics Group, Physics Department, Universidad de los Andes, Bogot\'{a} 111711, Colombia}
 
\author{Alonso Botero}%
\email{abotero@uniandes.edu.co}
\affiliation{Quantum Optics Group, Physics Department, Universidad de los Andes, Bogot\'{a} 111711, Colombia}




\date{\today}

\begin{abstract}
We present the asymptotic analysis of the quantum two-port interferometer in the $n \rightarrow \infty$ limit of $n$ partially indistinguishable photons. Using the unitary-unitary duality between port and inner-mode degrees of freedom, the probability distribution of output port counts can be decomposed as a sum of contributions from independent channels, each associated to a spin-$j$ representation of $SU(2)$ and, in this context, to $2 j$ effectively indistinguishable photons in the channel. Our main result is that the asymptotic output distribution is dominated by the $O(\sqrt{n})$ channels around a certain $j^*$  that depends on the degree of indistinguishability. The  asymptotic form is essentially the doubly-humped semi-classical envelope of the distribution that would arise from $2 j^*$ indistinguishable photons, and which reproduces the corresponding classical intensity distribution.
\end{abstract}

\maketitle



 The concept of interference is fundamental to our understanding of light, particularly in its classical wave-like manifestations, as demonstrated by the Michelson interferometer
 \cite{guenther_modern_2015}.
However, with the advent of photon-pair sources, a more nuanced type of interference has become accessible, namely the quantum interference of multiphotonic probability amplitudes \cite{feynman_feynman_2011} as exemplified by the Shih-Alley-Hong-ou-Mandel (HOM) experiment \cite{hong_measurement_1987,*shih_new_1988, *legero_time-resolved_2003, *bouchard_two-photon_2021}. The Michelson and HOM interferometers can be respectively understood as the macroscopic and microscopic limits of the standard scenario of two independent light sources interfering via a beam splitter \cite{campos_quantum-mechanical_1989}\cite{prasad_quantum_1987,*luis_quantum_1995, *fearn_quantum_1987}. A natural question then arises: what is the interplay between the classical and quantum notions of interference as the number $n$ of impinging photons scales from the two-photon HOM limit  to the macroscopic many-photon Michelson limit?
 
This connection between the granular and macroscopic behaviors of light at the beam splitter has been the subject of considerable interest \cite{campos_quantum-mechanical_1989,laloe_quantum_2012,tichy_many-particle_2012, tichy_interference_2014}. The best understood case is the extension of the HOM scenario to many perfectly indistinguishable photons; that is, identical photons in all degrees of freedom that are ancillary to the path (frequency, polarization, momentum, etc.), which henceforth will be named \emph{inner-modes}. Here, the problem can be framed in the language of angular momentum \cite{campos_quantum-mechanical_1989}, where the probability amplitudes are given by elements of the spin-$j$ representation matrix of the $SU(2)$ beam splitter  transformation. In the large $n$ limit,  a coexistence of classical and quantum interference effects \cite{campos_quantum-mechanical_1989,laloe_quantum_2012} is found: the probability distribution of output port counts shows rapid oscillations due to multiphotonic interference, modulated by an envelope corresponding to the classical distribution of output intensities from two interfering beams with a random relative phase. 

The case of the HOM setting with many partially indistinguishable photons is less understood. One approach is that of references \cite{tichy_many-particle_2012,*tichy_interference_2014,*tichy_extending_2017} where the resulting statistics are decomposed in terms of ``distinguishability types" labeled by the conserved occupation numbers in an orthogonal inner mode basis. More recent approaches to dealing with the problem of partial indistinguishability in multiphoton interference have emphasized a decomposition based on the so-called unitary-unitary  duality, where another good quantum number emerges labelling  $SU(N)$ irreducible representations \cite{stanisic_discriminating_2018,spivak_generalized_2022,tillmann_generalized_2015}. In the two-port ($N=2$) case, this number corresponds to a spin-$j$ common to path and inner mode degrees of freedom. Although these approaches can be extended to large numbers of photons, the asymptotic analysis of the output distribution in the HOM setting with partially indistinguishable photons has not been presented yet. 

\begin{figure}
    \centering
    \includegraphics[width = 0.45\textwidth]{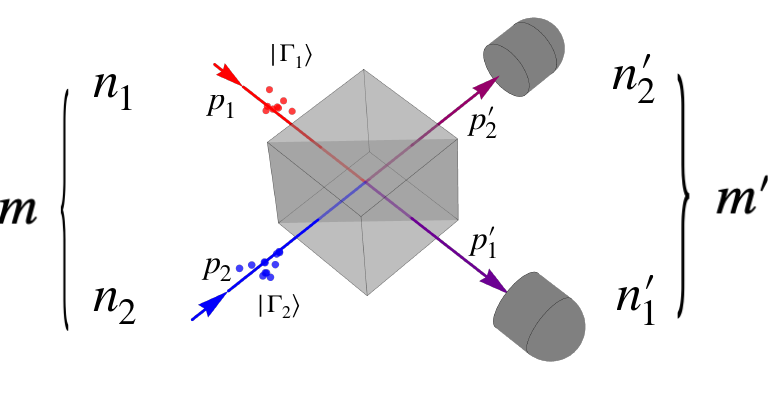}
    \caption{Schematic of the quantum two-port interferometer with partially indistinguishable photons. $m$ and $m'$ measure the input and output port imbalance respectively.}
    \label{fig:1}
\end{figure}

In this work we present this asymptotic analysis of the HOM multiphoton interferometer with partially indistinguishable photons using the decomposition based  on  unitary-unitary duality between port and inner-modes degrees of freedom. This decomposition makes evident how  entanglement between the two degrees of freedom is the decoherence mechanism underlying the loss of visibility of multiphoton interference. We show that  in the asymptotic limit, the output statistics are dominated by the interference of a well-defined fraction of effectively indistinguishable photons,  which precisely determines the correspondence between classical and quantum interference phenomena in the partially indistinguishable case.

The setting is illustrated in Fig. \ref{fig:1}: a total of $n=n_1+n_2$ photons are fed to the two input ports of a beam splitter (BS) in an initial state $\ket{\psi}_{in} \propto  (a_{p_1,\Gamma_1}^{\dagger})^{n_1}(a_{p_2,\Gamma_2}^{\dagger})^{n_2}\ket{0}$ of $n_i$ photons entering port $p_i$ with the same inner mode function $\Gamma_i$ ($i \in \{1,2\}$). The inner mode function overlap is described by the \emph{indistinguishability parameter} 
\begin{equation}
    \eta = \abs{\bra{\Gamma_1}\ket{\Gamma_2}}, 
    \label{eqn:second}
\end{equation}
 where the limits $\eta = 0$ and $1$  correspond to completely distinguishable  and  completely indistinguishable photons, respectively. Our aim is  the probability distribution of  detector counts $(n_1',n_2')$  at the output ports $p_1'$ and $p_2'$, as a function of $\eta$ and $U_p$, the $SU(2)$ representation matrix of the BS. To this end, we define the \emph{input and output imbalances} $m = (n_1-n_2)/2$ and $m'= (n_1'-n_2')/2$ respectively, and quantify the output statistics in terms of the \emph{output imbalance distribution} (OID) $p(m'|m)$ of measuring an output imbalance $m'$ given an input imbalance $m$.

%

%


%
%
To compute the OID, it would at first seem convenient to work in a Fock basis $\ket{n_{11},n_{12},n_{21},n_{22}}$ labelled by   joint occupations  $n_{a b}$ of four  \emph{orthogonal} single-particle states  $\ket{a}_p\ket{b}_q$,  where  $a,b \in \{1,2\}$ and subscripts $p$ and $q$ refer to  port and inner-mode degrees of freedom. Indeed, when $\abs{\bra{\Gamma_1}\ket{\Gamma_2}}  =0$,  the input state for the setting of Fig. \ref{fig:1} is  one element of the basis, namely $\ket{\psi}_{in}=\ket{n_1,0,0,n_2}$ (i.e., $n_{11}=n_1, n_{22}=n_2 $). However, for arbitrary $\eta$ the expansion of $\ket{\psi}_{in}$  involves many terms with general joint occupations $n_{ab}$. Moreover, the $n_{ab}$ are not individually conserved by the evolution through the BS, with the  only obvious conserved combinations being the  unobserved inner-mode occupations   $\sum_{a}n_{ab}$. Hence, the Fock basis is ill-suited   as it provides no good (i.e., conserved) quantum numbers for the port degree of freedom.

Instead of the Fock basis, we propose an alternative basis that we term the  \emph{coordinated spin basis} (CSB), which implements  the unitary-unitary duality \cite{basile_dual_2020,*rowe_dual_2012} in this context  as described in Ref. \cite{stanisic_discriminating_2018}.    
 This basis is naturally adapted to the decomposition of the $n$-photon Hilbert space $\mathcal{H}_n$ as the symmetric subspace $\mathcal{H}_n = Sym(\mathcal{H}_p^{\otimes n}\otimes \mathcal{H}_q^{\otimes n})$ of the tensor product of the single-particle Hilbert space $\mathcal{H}_p \otimes \mathcal{H}_q \cong \mathbb{C}^2 \otimes  \mathbb{C}^2$. In this decomposition, the  BS transformation and the inner-mode non-orthogonality can be described by tensor products $U_p^{\otimes n}$ and $ B_q^{\otimes n}$ acting on 
 $\mathcal{H}_P := (\mathcal{H}_p)^{\otimes n}$ and $\mathcal{H}_Q := (\mathcal{H}_q)^{\otimes n}$ respectively, where $B_q$ is the non-unitary operator  mapping  orthogonal inner-mode basis elements $ \ket{b}_q$  to the  states $\ket{\Gamma_b}_q$:
 \begin{equation}
    B_q  = \ket{\Gamma_1}\bra{1}_q + \ket{\Gamma_2}\bra{2}_q .
 \end{equation}
  Using the spin-$1/2$ correspondence of $\mathbb{C}^2$, the analogs of the total angular momentum operators $J^2$ for $\mathcal{H}_P$  and $\mathcal{H}_Q$ commute with $U_p^{\otimes n}$ and $ B_q^{\otimes n}$, and hence give good quantum numbers $j_P$ and $j_Q$. 
 Furthermore, when restricted to $\mathcal{H}_n$, permutational symmetry requires that $j_P = j_Q$. Hence, the CSB basis elements can be labeled 
as $\ket{j, m_P, m_Q}$,  where  $j=j_P=j_Q$, and $m_P$ and $m_Q$ are  eigenvalues of the  $J_z$ operators for $\mathcal{H}_P$ and $\mathcal{H}_Q$. The latter can be interpreted in  this context as mode occupation imbalances with $2m_{P}= n_{11}+n_{12}-(n_{21}+n_{22})$ and $2m_{Q}= n_{11}+n_{21}-(n_{12}+n_{22})$.

\begin{figure}
    \centering
    \includegraphics[width = 0.45\textwidth]{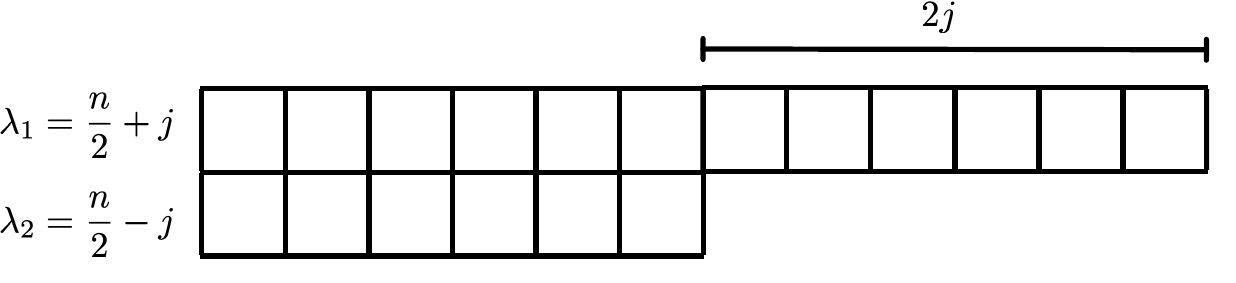}
    \caption{Young diagram representation of a two-row partition and its correspondence with the spin-$j$ IRRs of $SU(2)$.  }
    \label{fig:young}
\end{figure}
The CSB basis can  be understood from the so-called Schur-Weyl duality\cite{bacon_efficient_2006}, which entails the decomposition  
$(\mathbb{C}^2)^{\otimes n} \cong \bigoplus_{ \lambda} V_{\lambda} \otimes [\lambda] $, where $V_{\lambda}$ and $[\lambda]$ are respectively irreducible representations (IRR) of the linear group $GL(2,\mathbb{C})$ and the permutation group $S_n$, both labelled by the same Young diagram $\lambda$ \cite{fulton_young_1997,*hamermesh_group_1989}. The $V_{\lambda}$ IRR is a spin-$j$ representation  when  restricted to  $SU(2)$, with  Young diagram $\lambda_{j,n} = (\frac{n}{2}+j,\frac{n}{2}-j)$ (see  Fig. \ref{fig:young}). The dimension of $[\lambda_{j,n}]$, here denoted by $\gamma_{j,n}$, is 
\begin{equation}
    \gamma_{j,n} := \dim[\lambda_{j,n}] = \binom{n}{\frac{n}{2}+j} \frac{2j +1}{\frac{n}{2}+j +1}
    \label{eqn:multiplicity} \, ,
\end{equation}
which follows from the so-called \emph{Hook length formula} \cite{sagan_symmetric_2001}. 
The CSB  elements are    $S_n$-invariant states  from coupled    Schur-Weyl bases of $\mathcal{H}_P$ and $ \mathcal{H}_Q$. Denoting these  by $\ket{j,m_P,\tau}_P$ and $\ket{j,m_Q,\tau}_Q$, where $\tau$ labels  elements of a real basis of $[\lambda_{j,n}]$, the CSB elements  are maximally  entangled states between the $[\lambda_{j,n}]$ sectors:
\begin{equation}
    \ket{j,m_P,m_Q} = \frac{1}{\sqrt{\gamma_{j,n}}}\sum_{\tau} \ket{j,m_P,\tau}_P \otimes \ket{j,m_Q,\tau}_Q \, .
\end{equation}

In general, the transformation coefficients between the Fock and CSB bases can be complicated. However, one can show that the Fock state $\ket{n_1,0,0,n_2}$ admits a particularly simple  CSB basis expansion given by:
\begin{equation}
   \ket{n_1,0,0,n_2} =   \sum_{n/2 \geq j \geq m} \sqrt{\frac{\gamma_{j,n}}{\binom{n}{n_1}}}  \ket{j,m,m},
   \label{eqn:fockjjexp}
\end{equation}
with $m = m_P = m_Q = (n_1-n_2)/2$. This expansion is sufficient  for the computation of the OID, since the input and output states for our setting can be written as 
\begin{eqnarray}
  \ket{\psi}_{in} & = & \mathds{1}_P \otimes B_q^{\otimes n}\ket{n_1,0,0,n_2} \\ 
 \ket{\psi}_{out} & =  & U_p^{\otimes n} \otimes B_q^{\otimes n} \ket{n_1,0,0,n_2} \,,
 \label{eqn:vect} 
 \end{eqnarray}
and  the tensorized operators $U_p^{\otimes n}$ and $B_q^{\otimes n}$  act block-diagonally on the spin-$j$ sectors of  expansion \eqref{eqn:fockjjexp}. We  use the term \emph{channels} to refer to these sectors. The channel expansion for the output state is then
\begin{equation}
    \ket{\psi}_{out} =  \sum_{j\geq |m|} \sqrt{\frac{\gamma_{j,n}}{\binom{n}{n_1}}}\mathcal{D}^{j,n}(U_p)\otimes \mathcal{D}^{j,n}(B_q) \ket{j,m,m},
\label{eqn:outmultis}
\end{equation}
 where  $\mathcal{D}^{j,n}(g)$ are the spin-$j$ representation  matrices for $g \in GL(2,\mathbb{C})$. The OID is then $p(m'|m)= \left| \Pi_{m'} \ket{\psi}_{out} \right|^2 
    $, where  $\Pi_{m'} = \sum_{j,m_Q} \ket{j,m',m_Q}\bra{j,m',m_Q}$ projects onto the subspace of  port imbalance $m'$.    From  Eq. \eqref{eqn:outmultis}, one finally obtains  the OID channel expansion:
\begin{equation}
p(m'|m) = \sum_{|m| \leq j \leq n/2}  p(j|m)p(m'|j,m), 
 \label{eqn:prob}
\end{equation}
where $p(j|m)$, henceforth the \emph{channel probability}, is
\begin{equation}
    p(j|m) = \frac{\gamma_{j,n}}{\binom{n}{\frac{n}{2}+m}}\mathcal{D}^{j,n}_{m,m}(B_q^\dagger B_q ),
    \label{eqn:pj}
\end{equation}
and $p(m'|j,m)$, henceforth the \emph{channel OID}, is 
\begin{equation}
p(m'|j,m)=|\mathcal{D}^{j,n}_{m',m}({U_p})|^2
    \label{eqn:pm'}
\end{equation}
(for $j<\max(|m|,|m'|)$, $\mathcal{D}^{j,n}_{m',m}(g):=0$). Therefore, the OID is as a mixture of spin-$j$ channel contributions $p(m'|j,m)$  weighted by the channel probabilities $p(j|m)$. 
 
 Expressions \eqref{eqn:pj} and  \eqref{eqn:pm'} can be cast in terms of more familiar functions. The matrix element $\mathcal{D}^{j,n}_{m,m}(B_q^\dagger B_q )$ in Eq. \eqref{eqn:pj} can be written in terms of Gauss hypergeometric functions as
\begin{equation}    \mathcal{D}^{j,n}_{m,m}(B_q^\dagger B_q ) =   |1-\eta^2|^{\frac{n}{2}-j}   {}_2F_1(-j-m,-j+m;1;\eta^2).
     \label{eqn:hyper}
\end{equation}
Similarly, we can replace $\mathcal{D}^{j,n}_{m',m}$ in Eq. \eqref{eqn:pm'}  by Wigner's small $d$-matrix $d_{m'm}^j(\theta)$ when $U_p$ is written in Euler-angle form $U_p = e^{-i \phi \sigma_z/2}e^{-i \theta \sigma_y/2}e^{-i \psi \sigma_z/2}$. Thus,  the OID can be parameterized by  the indistinguishability $\eta$ and the reflectivity angle  $\theta$, where the channel probability only depends on $\eta$ and  the channel OID only depends on $\theta$. 

To illustrate the different terms of the OID in Eq.  \ref{eqn:prob}, we look at the standard HOM setting. In this case, $n=2$, $m=0$ and $j = 0$ or $1$, with no multiplicities ($\gamma_{j,2}=1$). The relevant   CSB basis elements, $\ket{0,0,0}$ and $\ket{1,0,0}$ are linear combinations of the product basis states $\ket{12}$, $\ket{21}$ in $\mathcal{H}_P=\mathcal{H}_p^{\otimes 2}$ and $\mathcal{H}_Q=\mathcal{H}_q^{\otimes 2}$:
\begin{equation}  
\left. \begin{array}{c} \ket{1,0,0}   \\ \ket{0,0,0} \end{array} \right\} = \ket{\pm}_P \ket{\pm}_Q\, , \qquad \ket{\pm} := \frac{ \ket{12} \pm \ket{21}}{\sqrt{2}}\, ,
\end{equation}
which are paired singlets ($-$) and  paired $m=0$ triplets ($+$). The matrix $\mathcal{D}^{j,n}(B_q) $ in Eq. \eqref{eqn:outmultis} in this case represents the action of $B_q^{\otimes 2}$ on $\mathcal{H}_Q$,  replacing  the inner mode states $\ket{a}_q$  for $\ket{\Gamma_{a}}_q$ and thus yielding channel probabilities $p(j|m)$:  
\begin{equation}
\left.\begin{array}{c} p(1|0)  \\ p(0|0) \end{array} \right\} = \frac{1}{2}\left| \frac{\ket{\Gamma_1 \Gamma_2} \pm \ket{\Gamma_2 \Gamma_1} }{\sqrt{2}}   \right|^2 = \frac{1 \pm \eta^2}{2} \, .
\end{equation}
In turn, $\mathcal{D}^{j,n}(U_p) $  here describes the action of $U_p^{\otimes 2}$ on  $\mathcal{H}_P$, leaving the singlet $\ket{-}_P$ invariant but transforming  $\ket{+}_P$  into the linear combination of triplet states: 
\begin{equation}
 U_p^{\otimes 2}\ket{+}_P= d^{1}_{0,0}(\theta) \ket{+}_P + d^{1}_{1,0}(\theta)\ket{11}_P + d^{1}_{-10}(\theta)\ket{22}_P  \, , 
\end{equation}
where $d^{1}_{0,0}(\theta) = \cos\theta$ and $d^{1}_{\pm1,0}  (\theta) = \mp \sin\theta/\sqrt{2}$. Thus,  the non-zero
channel OIDs $p(m'|j,m)$ are
\begin{equation}
p(0|0,0) = 1, \ \ p(\pm1|1,0) = \frac{1}{2} \sin^2\theta, \ \ p(0|1,0) = \cos^2\theta.
\end{equation}
The OID therefore involves two channels: 1) the spin-$1$ channel, with probability $(1+\eta^2)/2$, in which   photons enter  in a symmetric state in $\mathcal{H}_P$,  hence behaving at the BS as perfectly indistinguishable photons; and 2) the  spin-$0$ channel, with probability  $(1- \eta^2)/2$, where the photons form a $\mathcal{H}_P$ singlet  that passes  unaltered through the BS,  sending the photons to different output ports. When $\eta=1$, the $j=0$ channel is suppressed and the OID reproduces the HOM dip $p(0|0)=0$ for $\theta= \frac{\pi}{2}$. When $\eta=0$ both channels are equally weighted and the OID is  the distribution for perfectly distinguishable photons. These limits correlate with the   entanglement between $\mathcal{H}_P$ and $\mathcal{H}_Q$ in the  state $\ket{\psi}_{in}$, which is separable when $\eta=1$ and maximally entangled for $\eta=0$.

The case $n>2$ involves more channels.  When $\eta=1$, the only channel is $j=n/2$, the symmetric  channel with $PQ$-separable state. As $\eta$ is decreased,  $j<n/2$ channels are activated and  the state becomes $PQ$-entangled; finally, when $\eta=0$, the initial state  is  $\ket{n_1,0,0,n_2}$, which  maximally entangles the   $J_z=m$ sectors of  $\mathcal{H}_P$ and $\mathcal{H}_Q$.  To  interpret  the $j< n/2$ channels, we note from Eq. \eqref{eqn:pm'} and  $\det(U_p) =1$, that  $p(m'|j,m)$ is  the same as when $\eta =1$ and $n=2 j$. 
Thus, each term in Eq. \eqref{eqn:prob}  can be viewed as the contribution from  $2j$ perfectly indistinguishable photons.  These photons can be associated with the last $2j$ boxes in the first row of the Young diagram (Fig. \ref{fig:young}) for the partition $\lambda_{j,n}$.  The remaining  photons can  be associated to the scalar representation $V_{(0,n-2j)}$, corresponding to the first $n/2-j$ columns in the Young diagram, which  is proportional to a tensor product of $n/2 -j$ singlets. One can then view the spin-$j$ channel as  one where $2j$ photons behave as perfectly indistinguishable particles, while the remaining   photons pair up into $n/2 - j$ singlets,  passing unaltered through the BS. 

\begin{figure*}
    \centering
    \includegraphics[width =0.95\textwidth]{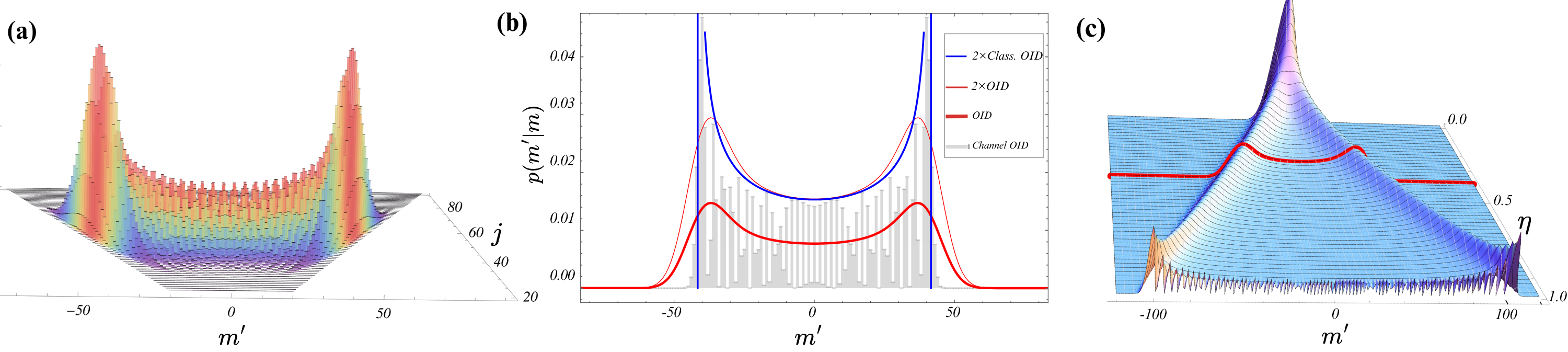}
    \caption{(a) Joint probability $p(j,m'|m) = p(j|m)p(m'|j,m)$, the summand of Eq. \eqref{eqn:prob} for the OID, as a function of $j$ and $m'$ for $n=240$, $m=0$, $\theta = \frac{\pi}{3}$, $\eta = 0.4$, showing concentration around $j^*=48$. (b) Comparison between the resulting OID $p(m'|m)$, the dominant channel OID $p(m'|j^*,m)$ and the classical OID (see Eq. \eqref{eqn:classOIDeq}) for the same parameters as in (a). The classical and quantum OID's are shown scaled by a  factor of $2$ to compare with the envelope of $p(m'|j^*,m)$. (c) OID as a function of $\eta$ and $m'$ with $n$, $m$ and $\theta$ as in (a) and (b), with the case for $\eta = 0.4$ indicated in red. }
    \label{fig:3}
\end{figure*}

To understand the large-$n$ behavior of the OID, we analyze the asymptotics of the channel probabillity $p(j|m)$ and the channel OID $p(m'|j,m)$. For the latter, we write $p(m'|j,m)=|d^{j}_{m'm}(\theta)|^2$ and following ref. \cite{braun_semiclassics_1996}, we use the fact that the Wigner small-$d$ matrix elements admit a  WKB approximation with an oscillatory ``classical'' regime and an exponentially decaying  regime determined by the sign of 
\begin{equation}
    R^j_{m'm} = (J^2- m^2)\sin^2\theta - (m'- m \cos \theta)^2
\end{equation}
where $J = \frac{1}{2}+j$ and  $R^j_{m'm} >0 $ corresponds to the classical region. In this region  $d^j_{m'm}$    takes the WKB form 
\begin{equation}
    d^j_{m'm}(\theta) = \frac{(-1)^j\sqrt{2/\pi}}{|R^j_{m'm}|^{1/4}}\cos(J S^j_{m'm} - \frac{\pi}{4}),
    \label{eqn:wkb}
\end{equation}
where  $S^j_{m'm}$ and  $R^j_{m'm}$ are related by $J \frac{\partial^2 S}{\partial m\partial m'}  = 1/\sqrt{R}$. Thus, as a function of $m'$, $p(m'|j,m)$ shows oscillations with $O(1)$ fringe separation at the center; as illustrated by the gray histogram in Fig. \ref{fig:3} (b) these oscillations
are modulated by an envelope $\propto (R^j_{m'm})^{-1/2}$ with its minimum at $\langle m'\rangle =m \cos\theta$, and peaks around the classical turning points
\begin{equation}
m'_\pm = m \cos \theta \pm \sqrt{J^2-m^2}\sin \theta.
\end{equation}
Thus $p(m'|j,m)$ has an $O(j)$ width, in stark contrast with the $O(\sqrt{j})$ width expected from  $2 j$ independent events. 

Proceeding with the   channel probability asymptotics, we find that as $n \rightarrow \infty$ with $j$ and $m$ growing proportionately, the distribution $p(j|m)$ becomes sharply peaked  around a certain value $j^*$ that depends on $\eta$ and $m$, with an $O(\sqrt{n})$  width. Introducing  scaled variables $\jb := j/n$ and $\mb :=m/n$, we find that $p(j|m)$ follows a large deviations principle $p(j|m) \sim e^{-nr(\jb)}$ \cite{touchette_basic_2012,sornette_critical_2000} and therefore concentrates at the value of $\jb$ where rate function $r(\jb) = -\lim_{n\rightarrow \infty} \frac{1}{n}\ln p(n\jb|n \mb)$ attains its minimum $\
\jb^* \in [1/2,\mb]$, which is given by
\begin{equation}
    \jb^* =  \sqrt{  \frac{\eta^2}{4} + (1- \eta^2) \mb^2 }. 
    \label{eqn:jjjj}
\end{equation}
  As expected, $j^* = n\jb^*$ is $n/2$ when $\eta=1$; the value $j^* = |m|$ for $\eta=0$ is explained from the fact that $2 |m|$ indistinguishable photons remain at one of the ports when the maximum number $\min(n_1,n_2)$ of singlets has been exhausted.

Combining  channel probability and channel OID asymptotics, together with our previous interpretation of the spin-$j$ channels, we arrive at a compelling picture to explain the OID asymptotic behavior.
From the concentration of measure of $p(j|m)$, only channels with  $|\jb - \jb^*| \sim \Delta j =  O(n^{-1/2})$  contribute significantly to the sum in Eq. \eqref{eqn:prob}. The fact that $\Delta j/j^* \sim O(n^{-1/2}) $ implies that when $n \rightarrow \infty$,  a sharply-defined fraction  $ 2 \jb^*$ of the  photons  behave  as indistinguishable particles, with the remaining fraction $1-2 \jb^*$ evenly distributed between the two output ports as singlets. 

The shape of the OID is obtained from the channel OIDs  with $|\jb - \jb^*| \sim \Delta j $, which in terms of scaled variables $\mb$,  share the same envelope, up to $O(n^{-1/2})$ corrections. However,  for adjacent values of $j$, the fringe patterns in $p(m'|j,m)$ show relative displacements of the order of half a fringe. Therefore, as illustrated in Fig. \ref{fig:3} (a) and (b), the fringe patterns are ``washed out" in the channel average and only appear when $\eta \sim 1$. Consequently, in the large-$n$ limit, the resulting shape of  $p(m'| m)$ is essentially the doubly-humped shape of the \emph{envelope} of $p(m'|j^*,m)$. In particular, in the classical region
\begin{equation}
p(m'| m) \propto \frac{1}{\sqrt{\eta^2 \left( \frac{1}{4} - \mb^2\right)\sin^2 \theta  - (\mb'- \mb \cos \theta )^2 }},
\label{eqn:classOIDeq}
\end{equation}
as depicted in Fig. \ref{fig:3} (b).

Similarly to the indistinguishable case\cite{campos_quantum-mechanical_1989,laloe_quantum_2012}, the OID $p(m'|m)$  is in direct correspondence with the classical OID of the output intensities from two  beams  with random relative phases impinging on the BS, but now exhibiting classical partial indistinguishability, say due to non-parallel polarizations.   To see this, suppose two   waves of the same frequency impinge on the BS, with electric field amplitudes  described by  vectors
$\boldsymbol{E}_1 \propto \sqrt{n_1}\boldsymbol{\varepsilon}_1$, and $\boldsymbol{E}_2 \propto \sqrt{n_2} \boldsymbol{\varepsilon}_2 e^{i \phi}$, where the $n_a$ are the intensities,  $\boldsymbol{\varepsilon}_a$ are normalized polarization vectors, and $\phi$ the relative phase. After passing through the BS, the output fields satisfy  
$
\boldsymbol{E}_i' = \sum_j U_{i j} \mathcal{R}_{ij} \boldsymbol{E}_j
$
where the $\mathcal{R}_{ij}$ are polarization rotation matrices  preserving the  inner product $\boldsymbol{\varepsilon_1}^* \cdot \boldsymbol{\varepsilon_2}$.
 The outgoing  intensities   are then given by $n_a'= |\boldsymbol{E}_a'|^2$, and the output imbalance $m'= (n_1'-n_2')/2$,  as a function of the relative phase, is
\begin{equation}
m'(\phi) = m \cos \theta + \eta \sqrt{ \frac{n^2}{4} - m^2  } \sin \theta \cos (\phi - \chi),
 \label{classimbal}
\end{equation}
where $n$ is the total intensity,   $\chi = \arg \boldsymbol{\varepsilon_1}^* \cdot \boldsymbol{\varepsilon_2}$, and $\eta := |\boldsymbol{\varepsilon_1}^* \cdot \boldsymbol{\varepsilon_2}|$. Thus, as  $\phi$ is varied, the classical output imbalance oscillates about a mean value $\langle m'\rangle = m \cos \theta$ with  amplitude  $\eta \sqrt{ \frac{n^2}{4} - m^2  }\, \sin \theta$. If $\phi$ is now uniformly randomized, the resulting probability distribution for $m'$ is exactly the right hand side of Eq. \eqref{eqn:classOIDeq}, with a proportionality constant of $n/\pi$. Therefore, the  OID is centered at the classical mean imbalance  and  is peaked at the  minimum and maximal values of $m'(\phi)$. Beyond the classical turning points, the exponential fall-off of the OID may differ somewhat from that of the  dominant channel OID. The explanation can be  appreciated from Fig. \ref{fig:3}(a): in the $(m,j)$ plane, the turning-point positions for  $|j -j^*| \sim \Delta j$ vary approximately linearly, with their height modulated by the  Gaussian shape of $p(j|m)$. Thus these turning-points  present a rotated Gaussian profile that distort the tails of the OID after marginalizing to $p(m'|m)$. 

The  $\eta$ dependence  of  the OID is illustrated in  Fig. \ref{fig:3} (c). The maximum width is obtained at  $\eta=1$, where the OID shows its full WKB form with interference fringes in the classical region. As $\eta$ decreases, the fringes disappear and the peak separation  decreases linearly with $\eta$ until the peaks coalesce into a single one, which  corresponds classically to the loss of interference for orthogonal  polarizations. In the quantum setting, the peak coalescence  signals a transition from  a multiphotonic correlated scenario to an independent photon regime. 
The OID for  $\eta=0$ can be explained in terms of $n$ i.i.d. samples of a classical binary symmetric channel (BSC) \cite{mackay_information_2005} with error probability  $f =\sin^2\frac{\theta}{2}$. The resulting  OID in this case is centered at  $\langle m' \rangle = m \cos \theta$ with variance
$
\langle \Delta m'^2 \rangle = \frac{n}{4} \sin^2 \theta \, $
and  is approximately Gaussian for large $n$. Hence, one expects a transition  when the $\propto n\eta$ width of the quantum OID  becomes comparable with the $\propto \sqrt{n}$ width of the BSC distribution. This suggests a critical value,  $\eta_c$, scaling as $\eta_c \sim n^{-1/2}$, a scaling that we have observed numerically.

Finally, in response to the question posed in the introduction, we have shown a coexistence between the classical wave-like and the quantum multiphotonic notions of interference in the asymptotic analysis of the quantum beam splitter. While the asymptotic form of the OID reproduces its classical counterpart, its granular interpretation comes from the multiphotonic interference of a well-defined fraction of perfectly indistinguishable photons, which is determined by combinatorial functions and the degree of indistinguishability. It would be interesting to further investigate the extension of our analysis to the case of more than two ports. 

\begin{acknowledgements}
 A.V. and A.B. acknowledge support from the Faculty of Sciences at Universidad de los Andes. 
\end{acknowledgements}


\bibliography{main}

\end{document}


\title{Supplemental Material for ``The quantum beam splitter with many partially indistinguishable photons: multiphotonic interference and asymptotic classical correspondence'' }
\maketitle



\section{Preliminaries}
This section provides some required background on the combinatorial and group theoretical techniques techniques that are necessary for the derivation of the main results of the paper.

\subsection{Product basis expansion of the $n$-boson Fock basis}
Consider  a system of $n$ identical particles with single-particle Hilbert space $\mathcal{H}_1$ of finite dimension $d$. Given some standard basis for $\mathcal{H}_1$ with basis elements $ \ket{i}$,  the product basis for the $n$-particle Hilbert space $\mathcal{H}_1^{\otimes n}$ can be labeled by  sequences $s  = (i_1 i_2 \ldots i_n) \in \{ 1, \ldots d\}^n$ of index values:
\begin{equation}
    \ket{s} = \ket{i_1i_2,...i_n} = \ket{i_1}\otimes\ket{i_2}\otimes ... \otimes \ket{i_n}.
\end{equation}
 If the particles are indistinguishable bosons, the $n$-particle Hilbert space is further restricted to be the symmetric subspace of $\mathcal{H}_1^{\otimes n}$, 
$$
\mathcal{H}_n = \mathrm{Sym}_n \mathcal{H}_1.
$$
A natural basis for $\mathcal{H}_n$ can be obtained from projecting product basis states to the symmetric subspace using the projector operator
\begin{equation}
\Pi_{\mathrm{Sym}} = \frac{1}{n!}\sum_{\pi \in S_n} U(\pi),
\end{equation}
where $\pi$ runs over all permutations of $n$ symbols and $U(\pi)$ acts on a product basis state
as
\begin{equation}
  U(\pi)\ket{s} = \ket{\pi s} = \ket{ i_{\pi^{-1}(1)} \ldots i_{\pi^{-1}(n)} }.
\end{equation}
From the definition of $\Pi_{\mathrm{Sym}}$, we see that two product basis elements labeled by sequences that differ from each other by a permutation project to the same state. Hence, the possible states obtained from   projecting the product  basis states  are in one-to-one correspondence with the equivalence classes of sequences in $\{ 1, \ldots d\}^n$  under the equivalence relation  $s \sim s'$ iff $s'= \pi s$ for some $\pi \in S_n$.  Each class is composed of sequences of the same \emph{type}, that is, sequences with the same set of 
occupation numbers $ (n_1, \ldots n_d)$, where $n_i$ counts the number of occurrences of the value $i$ in the sequence.
Hence, images of the product  basis elements can be labelled by   the set of occupation numbers $ (n_1, \ldots n_d)$ and, once normalized, are given by the superposition 
\begin{equation}
\ket{n_1, n_2, \ldots n_d } = \left( \frac{ n!}{n_1! n_2 ! \ldots n_d ! } \right)^{-1/2} \sum_{s \sim (n_1, n_2, \ldots n_d) } \ket{s}
\label{Fockexp}
\end{equation}
where $s \sim (n_1, n_2, \ldots n_d)$ means that $s$ is a sequence of type $(n_1, \ldots n_d)$, i.e.,  equivalent under permutations to the sequence 
$$
(\underbrace{11\ldots1}_{n_1 \mathrm{ \ times}} \, \underbrace{22\ldots2}_{n_2 \mathrm{ \ times}}\, \ldots  \underbrace{dd\ldots d}_{n_d \mathrm{ \ times}}).
$$
The set  of  all basis  elements given by Eq.  \eqref{Fockexp}, with $\sum_i n_i =n$, spans the symmetric space $\mathcal{H}_n$ of $n$ bosons \cite{harrow_church_2013}. Expansion \eqref{Fockexp} is the first-quantization representation of the  second-quantization $n$-particle Fock basis states $\ket{ n_1, \ldots n_d}$ for bosons.

\subsection{$GL(d, \mathbb{C})$ symmetric representations and $SU(2)$ IRRs as bosonic Hilbert spaces}

 Since $\mathcal{H}_1$ is $d$-dimensional, it is the defining representation of the group $GL(d, \mathbb{C})$ of $d \times d$ complex invertible matrices. This group acts naturally on $\mathcal{H}_1^{\otimes n}$ by the representation $\rho(g)$ corresponding to the application of the same group element to every copy of $\mathcal{H}_1$, that is
 \begin{equation}
  \rho(g)   \ket{\phi_1}\otimes\ket{\phi_2}\otimes ... \otimes \ket{\phi_n} = g^{\otimes n} \ket{\phi_1}\otimes\ket{\phi_2}\otimes ... \otimes \ket{\phi_n} = g\ket{\phi_1}\otimes g\ket{\phi_2}\otimes ... \otimes g\ket{\phi_n},
 \end{equation}
 for any $g \in GL(d, \mathbb{C})$. Furthermore $GL(d, \mathbb{C})$ acts irreducibly when restricted to the symmetric subspace. Hence, the bosonic spaces $\mathcal{H}_n$ are  carrier spaces of the so-called \emph{symmetric IRR}s of $GL(d, \mathbb{C})$, which are labelled by  Young diagrams of $n$ boxes with only one row, i.e., $\lambda= (n) $. 
  From expression \eqref{Fockexp}, the matrix elements of $\rho(g)$ in the Fock basis are given by
 \begin{equation}
     \bra{n_1',n_2',\ldots n_d'}\rho(g)\ket{n_1, n_2, \ldots n_d} =  \frac{ \left( \prod_{i=1}^{d} n_i ! n_i'! \right)^{1/2}  }{n!}  \sum_{s' \sim (n_1', n_2', \ldots n_d')}\sum_{s \sim (n_1, n_2, \ldots n_d) } \bra{s'} g^{\otimes n}  \ket{s}.
     \label{Dsuminseqs}
 \end{equation}
 Expanding $\bra{s'} g^{\otimes n}  \ket{s}$ as a product of matrix elements
 \begin{equation}
\bra{s'} g^{\otimes n}  \ket{s} = g_{i_1'i_1}g_{i_2'i_2}\ldots g_{i_n'i_n}
 \end{equation}
and collecting all monomials in the matrix elements $g_{ij}$, one obtains the explicit expression for the symmetric representation matrices as polynomial in the matrix elements $g_{ij}$
 \begin{equation}
     \bra{n_1',n_2',\ldots n_d'}\rho(g)\ket{n_1, n_2, \ldots n_d} =   \left( \prod_{i=1}^{d} n_i ! n_i'! \right)^{1/2}  \sum_{W}^*\prod_{ij} \frac{ g_{ij}^{W_{ij} }   }{ W_{ij}! },
     \label{SymDexp}
 \end{equation}
where the sum runs over all $d\times d$ matrices $W$ with non-negative integer-valued elements such that the rows sum up to the occupations $n_i'$ and the columns to the occupations $n_i$\cite{louck_unitary_2008}. 

We note that $\rho(g)$ is not generally a unitary IRR for   $g \in GL(d, \mathbb{C})$, but is  when restricted to $g \in U(d)$ where $U(d) \subset GL(d, \mathbb{C})$ is the group of $d\times d$  unitary matrices. In such a case, $g$ represents a basis change to a new orthonormal basis in the single-particle Hilbert space, and its action on $\mathcal{H}_n$ can be seen in second-quantized language as the transformation of creation/annhiliation operators
 \begin{equation}
     \rho^{\dagger}(g) a_i \rho(g) = g_{ij} a_j, \qquad  \rho^\dagger(g) a_i^\dagger \rho(g) = g_{ij}^* a_j^\dagger,
 \end{equation}
  which by unitarity of $g$ commutes with the total number operator $N=\sum_{i=}^d  a_i^\dagger a_i$.

When $d=2$, $n+1$-dimensional bosonic Hilbert spaces $\mathcal{H}_n$ are in one-to one correspondence with the $2j + 1$-dimensional spin-$j$ representations of $SU(2) \subset GL(2,\mathbb{C})$, for $j=n/2$.  Thinking of the index values $i=1$ and $i=2$ as respectively labeling spin-up ($m= 1/2)$ and spin-down  $(m= 1/2)$ states of the $j=1/2$ representation,  the Fock state $\ket{n_1,n_2}$ with $n_1+n_2 = n$ corresponds to the spin-$j$  angular momentum state $\ket{j,m}$ with
\begin{equation}
    j = \frac{n_1 + n_2}{2}, \qquad  m = \frac{n_1 - n_2}{2},
\end{equation}
or equivalently
\begin{equation}
    n_1  =  j+ m , \qquad  n_2 = j-m.
\end{equation}
Imposing the row and sum conditions for the  $ 2 \times 2 $ matrix $W$ in the specialization of  expansion \eqref{SymDexp} to $d=2$ leaves only one free element; letting this  element be $x := n_{12}$, we have
\begin{equation}
    W(x) = \left(\begin{array}{cc}j + m'- x & x \\m-m'+ x & j-m -x\end{array}\right), 
\end{equation}
and consequently expansion of the spin-$j$ representation matrices as  polynomials in the matrix elements of $g \in SU(2)$:
\begin{equation}
    D^j_{m',m}(g)= \sqrt{(j+m)!(j-m)!(j+m')!(j-m')!} \sum_x \frac{g_{11}^{j+m'-x}\, g_{12}^{x}\,  g_{21}^{m-m'+x}\,  g_{22}^{j-m-x}}{(j+m'-x)! x! (m-m'+x)! (j-m-x)! }
    \label{SU(2)polyexp}
\end{equation}
 where the sum runs over all values of $x$ for which the factorials in the summand are non-negative. Taking $g$ to be the matrix 
 \begin{equation}
    g = \left(\begin{array}{cc}\cos \frac{\theta}{2} & \sin \frac{\theta}{2} \\-\sin \frac{\theta}{2} & \cos \frac{\theta}{2}\end{array}\right), 
\end{equation}
and substituting into Eq. \eqref{SU(2)polyexp} gives the explicit expression for the Wigner small-$d$ matrices. 

\subsection{$GL(2,\mathbb{C})$  IRRs }
As discussed in the paper, the $2j +1$-dimensional  spin-$j$  representation spaces are also carrier spaces  of the $2j +1$-dimensional IRRs of $GL(2,\mathbb{CC})$, the group of $2 \times 2$ complex invertible matrices. These representations are labelled by two-row Young diagrams $\lambda = (\lambda_1, \lambda_2)$, with $\lambda_1 \geq \lambda_2$, and include the symmetric representations just discussed when $\lambda_2 =0$. Alternatively, we can label the IRRs using the pairs $(n,j)$ and the correspondence between $(\lambda_1,\lambda_2)$ and $(n,j)$ depicted in Fig. 2 of the paper, namely
\begin{equation}
    \lambda_1 = \frac{n}{2} + j,  \qquad  \lambda_2 = \frac{n}{2} - j, 
\end{equation}
or conversely,
\begin{equation}
    n = \lambda_1 + \lambda_2, \qquad  j = \frac{\lambda_1 - \lambda_2}{2}.
\end{equation}
 The  IRRs are obtained by a symmetrization-antisymmetrization algorithm  encoded in the Young diagram (see e.g., \cite{hamermesh_group_1989, balachandran_group_2010}). For a given pair $(n,j)$ the  elements of the representation matrices for $g \in GL(2,\mathbb{C})$ are given by the same polynomials $D^{j}(g)$ of the spin-$j$ representation matrices, as per  Eq. \eqref{SU(2)polyexp}, but multiplied by the determinant of $g$ to the power $\lambda_2 = \frac{n}{2} - j$. Labeling the  matrix elements by $\mathcal{D}^{j,n}_{m'm}(g)$, they are therefore given by 
 \begin{equation}
\mathcal{D}^{j,n}_{m'm}(g) = \det(g)^{\frac{n}{2} - j } \ D^{j}_{m'm}(g)
 \end{equation}
where $D^{j}_{m'm}(g)$ are as given in Eq. \eqref{SU(2)polyexp}. The matrix element $\mathcal{D}^{j,n}_{m'm}(g)$ admits an expansion as a polynomial in the matrix elements $g_{ij}$ similar to that of Eq. \eqref{SU(2)polyexp}. Using the latter expansion and multiplying with the expansion of $\det(g)^{\frac{n}{2} - j }$, one arrives at the expansion of the form
\begin{equation}
     \mathcal{D}_{m'm}^{j,n}(g) =  \sum_{W}^* C^{j,n}_{m'm}(W) \prod_{ij} \frac{n! }   { W_{ij}! }g_{ij}^{W_{ij} },
     \label{Louckexp}
 \end{equation}
where  the matrix $W$ is again a $2 \times 2$ integer-valued matrix with non-negative entries, but satisfying the condition that the rows sum to $\frac{n}{2} \pm m'$ and the columns to $\frac{n}{2} \pm m$, and $C^{j,n}_{m'm}(W)$ are matrix valued coefficients also known in the literature as Louck polynomials (\cite{chen_combinatorics_1998,louck_unitary_2008}). With a suitable parametrization of $W$, these polynomials can be expressed in terms of hypergeometric functions (see the appendix of \cite{botero_universal_2018}).

\subsection{Schur transform for $(\mathbb{C}^2)^{\otimes n}$ }

Schur-Weyl duality\cite{goodman_symmetry_2009} states that for single-particle Hilbert space $\mathcal{H}_1$, its $n$-fold tensor product  can be decomposed as the direct sum 
\begin{equation}
    \mathcal{H}_1^{\otimes n} = \oplus_{\lambda} V_\lambda \otimes [\lambda], 
\end{equation}
where $\lambda$ are partitions (i.e., Young diagrams) of $n$ of at most $d = \mathrm{dim} \mathcal{H}_1$ parts, and $V_\lambda$ and $[\lambda]$ are  respectively the carrier spaces for IRRs of  $GL(d, \mathbb{C})$  and the permutation group $S_n$, both labeled by $\lambda$ 
\cite{hamermesh_group_1989,balachandran_group_2010,fulton_representation_2013}. Thus,  $[\lambda]$ acts as the multiplicity space of the representation $V_\lambda$ and vice versa. In particular, the symmetric representation is the  $V_\lambda$ IRR for the one-row partition $(n)$, and is multiplicity-free given that $[(n)]$ is one-dimensional. 

For the case $d=2$, the basis adapted to the Schur-Weyl duality of $\mathcal{H}_1^{\otimes n}$ can be conveniently labelled by three quantum numbers: the spin-$j$ quantum number $j$, which together with $n$ gives the partition $\lambda_{j,n}=(\frac{n}{2}+j, \frac{n}{2}-j)$; the ``magnetic'' quantum number $m$  labeling the basis elements of the spin-$j$ IRR $V_{\lambda_{j,n}}$; and a third label $\tau$ for the multiplicity of the spin-$j$ representation, which acts as a label for the $S_n$ IRR $[ \lambda_{j,n} ]$ basis elements. The  elements can thus be written as
$$
\ket{ j, m, \tau}.
$$
The so-called \emph{Schur transform} \cite{bacon_efficient_2006} implements the basis change from the product basis $\ket{s}$ to a Schur-Weyl basis $|j,m ,\tau\rangle$ in a particularly transparent way. The idea is to build the basis progressively by performing angular momentum addition one spin-1/2 at a time, with the spin to be added at the $k$-th corresponding to the $k$-th element of the sequence $s$. The index $\tau$ therefore runs over all  admissiible sequences of intermediate angular momentum numbers up to the final angular momentum $j$:
\begin{equation}
  \tau =   (j_1 j_2,\ldots,j_{n-1},j_n),  \qquad j_{k+1} = |j_{k}\pm1/2|, \ \ j_0:=0, \ \ j_n := j
\end{equation}
The transformation matrix element from the product basis to the Schur-Weyl basis can then be written as a product of Clebsch-Gordan coefficients
\begin{equation}
\langle j, m , \tau | s \rangle = \prod_{k=1}^{n} \left\langle j_k, m_k \left| j_{k-1}, m_{k-1}; \frac{1}{2}, \frac{3}{2}-i_k \right. \right \rangle, \qquad m_k = m_{k-1}+ \frac{3}{2}-i_k,\ \  m_0:= 0
\end{equation}
where here we assume  that $i_k \in \{1,2\}$, as before. We note that from the reality of the Clebsch-Gordan coefficients, 
$$
\langle j, m , \tau | s \rangle = \langle j, m , \tau | s \rangle^* = \langle s | j, m , \tau \rangle,
$$
a property that will prove useful shortly. Another  consequence that follows from this construction is  the reality of the representation matrices of the permutations, that is, the matrices $S_{\tau',\tau}^{j,n}(\pi)$ such that
\begin{equation}
U(\pi) \ket{j, m, \tau} = \sum_{\tau'} S_{\tau',\tau}^{j,n}(\pi) \ket{j, m, \tau}.
\end{equation}

Several useful expansions involving the Schur transform are worth noting. First, In the case of binary sequences, the quantum number $m$ suffices to determine the type of the sequence according to
\begin{equation}
    n_1 = \frac{n}{2} + m, \ \ n_2 = \frac{n}{2} - m,
\end{equation}
so the projection onto the subspace of $(\mathbb{C}^2)^{\otimes n}$ with $J_z = m$ is the projector onto the subspace of sequences of type $(\frac{n}{2} + m \, ,  \frac{n}{2} - m)$:
\begin{equation}
\Pi_m = \sum_{j\geq|m|}\sum_{\tau=1}^{\gamma_{j,n}}\ket{j,m, \tau}\bra{j,m, \tau} \ \ \ = \sum_{s \sim (\frac{n}{2} + m \, , \frac{n}{2} - m) } \ket{s}\bra{s}.
\label{projm}
\end{equation}
Expanding one basis in terms of the other in the above expression, we obtain the relations
\begin{equation}
 \sum_{j\geq|m|}\sum_{\tau=1}^{\gamma_{j,n}}\langle s | j,m, \tau \rangle \langle j,m, \tau | s' \rangle =  \left\{ \begin{array}{cc} \delta_{s,s'}, & s \sim (\frac{n}{2} + m \, ,  \frac{n}{2} - m) \\
 0 , & s \nsim (\frac{n}{2} + m \, ,  \frac{n}{2} - m) ,
 \end{array}\right.
 \label{rels1}
\end{equation}
and 
\begin{equation}
 \sum_{s \sim (\frac{n}{2} + m \, , \frac{n}{2} - m) }
  \langle j'',m'', \tau'' | s \rangle\langle s | j',m', \tau' \rangle 
 =  \left\{ \begin{array}{lc} \delta_{j',j''} \delta_{m,m'}\delta_{m,m''}\delta_{\tau',\tau''}, & j'  \geq |m| \\
 0 , & j'  < |m| ,
 \end{array}\right.
 \label{rels2}
\end{equation}
as well as all associated relations obtained by interchanging any of the  $\langle j, m , \tau | s \rangle$ for $\langle s | j, m , \tau \rangle$. 

Finally, it is worth mentioning that by Schur's Lemma \cite{fulton_representation_2013}, there are no non-zero    permutationally invariant linear operators mapping one IRR $[\lambda]$ of $S_n$ to an inequivalent one $[\lambda']$, and within a given 
 IRR $[ \lambda ]$ of $S_n$, the only invariant operators are those that are multiples of the identity. This implies that the space of $S_n$ invariant operators in $(\mathbb{C}^2)^{\otimes n}$ is spanned by operators of the form
\begin{equation}
    X_{j,m,m'} = \sum_{\tau=1}^{\gamma_{j,n}}  \ket{j,m, \tau}\bra{j,m', \tau}, 
\end{equation}
which, when restricted to $V_{j,n}\otimes[\lambda_{j,n}]$ are equivalent to the operator 
$$
\ket{j,m}\bra{j,m'} \otimes \mathds{1}_{[\lambda_{j,n}]}.
$$ 
Since the matrix element $\bra{s}  X_{j,m,m'} \ket{s'}$ has the property that
\begin{equation}
   \bra{s}  X_{j,m,m'} \ket{s'}   =  \bra{\pi s}  X_{j,m,m'} \ket{\pi s'}, 
\label{xinv}
\end{equation}
it only depends on the equivalence class of sequence pairs $(s,s')$, under the equivalence relation $(s,s') \sim (s'', s''') \Leftrightarrow (s,s') = (\pi s'', \pi s''') $ for some $\pi \in S_n$. The invariant property of these equivalence classes are the joint  types of the  sequence pair $(s,s')$. These types are the occurences (or occupation numbers) $n_{ij}$ of the pair $(i,j)$ in the sequence of pairs $((i_1,i_1')(i_s,i_2')\ldots (i_n,i_n'))$ formed from pairing up the elements of the two sequences $s$ and $s'$. Remarkably, if we form the matrix $W(s,s')$ with its elements given by the joint weights $n_{ij}$ of $(s,s')$, the matrix element
$\bra{s}  X_{j,m,m'} \ket{s'}$ is given by 
\begin{equation}
    \bra{s}  X_{j,m,m'} \ket{s'} = {\gamma_{j,n}} C^{j,n}_{m m'}(W(s,s'))
    \label{Xmatel}
\end{equation}
where $C^{j,n}_{m m'}(W(s,s'))$ are the Louck polynomials in expansion \eqref{Louckexp}. This identity can be proved from the identity
\begin{equation}
\frac{1}{\gamma_{j,n}} \sum_\tau \bra{ j, m, \tau} g^{\otimes n} \ket{ j, m, \tau} = \mathcal{D}^{j,n}_{m,m'}(g)
\end{equation}
by expanding the left-hand side in the product basis, the right hand side using \eqref{Louckexp}, and after using Eq. \eqref{xinv}, comparing coefficients of the monomials in the matrix elements $g_{ij}$.

\section{Derivations of results involving the CSB basis}

In this section we use results from the previous section to derive results for the symmetric space $\mathcal{H}_n= Sym_n \mathcal{H}_1$ associated to a single-particle Hilbert space that is the tensor product of two two-dimesnional Hilbert spaces spaces $\mathcal{H}_p$ and $ \mathcal{H}_q $:
\begin{equation}
\mathcal{H}_1 = \mathcal{H}_p \otimes \mathcal{H}_q = \mathbb{C}^2 \otimes \mathbb{C}^2.
\end{equation}
We shall exploit the  fact that
\begin{equation}
    \mathcal{H}_1^{\otimes n} = (\mathcal{H}_p \otimes \mathcal{H}_q)^{\otimes n} \cong (\mathcal{H}_p)^{\otimes n} \otimes (\mathcal{H}_q)^{\otimes n},
\end{equation}
so states in $\mathcal{H}_1^{\otimes n}$ can be expanded in a basis that is a tensor product of the product bases for $\mathcal{H}_P := (\mathcal{H}_p)^{\otimes n}  $ and $\mathcal{H}_Q := (\mathcal{H}_q)^{\otimes n}  $:
\begin{equation}
    \ket{\psi} = \sum_{s,s'} \psi(s, s') \ket{s}_P \ket{s'}_Q'.
\end{equation}
Equivalently, transforming to the Schur-Weyl bases of 
$\mathcal{H}_P$ and $\mathcal{H}_Q$, the same state can be expanded as
\begin{equation}
    \ket{\psi} = \sum_{j,m,\tau, j',m',\tau'} \widetilde{\psi}(j,m,\tau, j',m', \tau'') \ket{j,m,\tau}_P \ket{j',m',\tau'}_Q'.
\label{psiSWexpand}
\end{equation}
Since the Schur transform coefficients $\langle j, m ,\tau| s \rangle$ are real, we can associate to $\ket{\psi}$ an operator $\mathcal{O}_\psi$ acting on $(\mathbb{C}^2)^{\otimes n}$ by
\begin{equation}
    \mathcal{O}_\psi := \sum_{s,s'} \psi(s, s') \ket{s} \bra{s'},
\end{equation}
and the coefficients $\widetilde{\psi}(j,m,\tau, j',m', \tau'')$  in  \eqref{psiSWexpand} can be understood as the matrix elements of $\mathcal{O}_\psi$ but in the Schur-Weyl basis:
\begin{equation}
\widetilde{\psi}(j,m,\tau, j',m', \tau'') = \bra{j,m,\tau} \mathcal{O}_\psi \ket{j',m',\tau'}.
\end{equation}
Furthermore, if we restrict ourselves to  operations on $\mathcal{H}_n$ of the form $\ c_i A_i \otimes B_i$ where 
$A_i$ and $B_i$ have \emph{real} representation matrix elements in either  basis (and hence in both),  we can move back and forth between representations using 
\begin{equation}
   \ket{\psi'}= \sum_i A_i \otimes B_i |\psi \rangle \leftrightarrow \mathcal{O}_{ \psi'} = \sum_i c_i A_i \mathcal{O}_\psi B_i^\dagger = \sum_i c_i A_i \mathcal{O}_\psi B_i^T 
\end{equation}
 where  $\leftrightarrow$ is the  linear extension of the bijection  $\ket{s}_P \ket{s'}_Q \leftrightarrow \ket{s} \bra{s'}$.
We can then exploit this bijection to connect properties of linear operators in $(\mathcal{C}^2)^{\otimes n}$ to properties of entangled states in $\mathcal{H}_P \otimes \mathcal{H}_Q$. 

\subsection{Derivation of Eq (4) }
The first derivation is that of the CSB basis of $\mathcal{H}_n= \mathrm{Sym}_N \mathcal{H}_p \otimes \mathcal{H}_q$ in terms of the Schur-Weyl bases of  $\mathcal{H}_P$ and $\mathcal{H}_Q$. We recall from subsection{...}, that the algebra of permutationally-invariant operators in $(\mathcal{C}^2)^{\otimes n}$ is spanned by the operators
\begin{equation}
    X_{j,m,m'} = \sum_{\tau=1}^{\gamma_{j,n}}  \ket{j,m, \tau}\bra{j,m', \tau}.
\end{equation}
Permutational invariance implies that
\begin{equation}
U(\pi) X_{j,m,m'} U(\pi)^\dagger = X_{j,m,m'}.
\end{equation}
Since $U(\pi)$ has real matrix elements in the Schur transform basis $\ket{j,m \tau}$, this implies that the under the the bijection  $\ket{s}_P \ket{s'}_Q \leftrightarrow \ket{s} \bra{s'}$, the entangled states
\begin{equation}
\widetilde{\ket{j,m,m'}} := \sum_{\tau} \ket{j,m, \tau} \ket{j, m',\tau} \leftrightarrow X_{j,m,m'}  
\end{equation}
are invariant under the same permutation applied to the $P$ and $Q$ spaces:
\begin{equation}
    U(\pi) \otimes U(\pi) \widetilde{\ket{j,m,m'}}= \widetilde{\ket{j,m,m'}},
\end{equation}
and hence are permutationally invariant under permutations of the different copies of $\mathcal{H}_p \otimes \mathcal{H}_q$. 
The  CSB basis elements are simply the states $\widetilde{\ket{j,m,m'}}$ when normalized.
Finally, since we know from Schur's lemma that no non-zero permutationally invariant operators exist connecting different $j$ sectors in  $(\mathcal{C}^2)^{\otimes n}$, 
 the bijection $\ket{s}_P \ket{s'}_Q \leftrightarrow \ket{s} \bra{s'}$ implies that no permutationally invariant states exist in $\mathcal{H}_P \otimes \mathcal{H}_Q$ involving different values of $j$ for $P$ and $Q$. Therefore, the CSB states span the permutationally symmetric subspace of $\mathcal{H}_1^{\otimes n}$.

\subsection{Derivation of Eq (5) }
A similar trick can be used to derive the expansion (5) of the Fock state $\ket{n_1,0,0,n_2}$ in terms of the CSB basis. For this we first turn to the first quantized representation of the Fock states $\ket{n_{11},n_{12},n_{21},n_{22}}$ in terms of a product basis of $\mathcal{H}_1^{\otimes n}$. Thinking of pairs $(a,b)$ with $a,b \in \{1, 2\}$ as symbols of a four-symbol alphabet, the Fock state can be written as 
\begin{equation}
    \ket{n_{11},n_{12},n_{21},n_{22}} = \sqrt{\frac{n_{11}! n_{12}! n_{21}! n_{22}!}{n!} }\sum_{S \sim (n_{11},n_{12},n_{21},n_{22}) } \ket{S} ,
\end{equation}
where $S$ runs over all sequences of pairs $(a,b)$ where the pair $(a,b)$ occurs $n_{ab}$ times. It is easier, however to think of $S$ as composed from a pair of binary sequences $(s, s') $, where $s = (a_1 a_2 \ldots a_n)$ and $s' = (b_1 b_2 \ldots b_n)$. When represented in this way, and decomposing  $\mathcal{H}_n $ as
\begin{equation}
    \mathcal{H}_n = \mathcal{H}_P \otimes \mathcal{H}_Q,
\end{equation}
we have 
\begin{equation}
    \ket{n_{11},n_{12},n_{21},n_{22}} = \sqrt{\frac{n_{11}! n_{12}! n_{21}! n_{22}!}{n!} }\sum_{(s, s') \sim (n_{11},n_{12},n_{21},n_{22}) } \ket{s}_P\ket{s'}_Q.
    \label{Fockgeninseq}
\end{equation}
The case  in which  $n_{11}=n_1$ and $n_{22}=n_2$ with $n_{12}=n_{21}=0$ corresponds to the situation where $s= s'$ and both sequences are binary sequences of type $(n_1,n_2)$. Hence,
we have 
\begin{equation}
    \ket{n_{1},0,0,n_{2}} = \frac{1}{\sqrt{\binom{n} {n_1} }}\sum_{s \sim (n_1,n_2) } \ket{s}_P\ket{s}_Q.
\label{Fockn1n2seq}
\end{equation}
Using the bijection $\ket{s}_P \ket{s'}_Q \leftrightarrow \ket{s} \bra{s'}$, the state maps to the projector operator onto the $J_z=m$ subspace of $(\mathbb{C}^2)^{\otimes n}$ and from \eqref{projm} we have the correspondence
\begin{equation}
 {\sqrt{\binom{n} {n_1} }} \ket{n_{1},0,0,n_{2}} \leftrightarrow   \Pi_m =   \sum_{j \geq |m|}\sum_{ \tau=1}^{\gamma_{j,n}}\ket{j,m ,\tau}\bra{j,m ,\tau}    \leftrightarrow   \sum_{j \geq |m|} \sum_{ \tau=1}^{\gamma_{j,n}}  \ket{j,m ,\tau}_P\ket{j,m ,\tau}_Q
\end{equation}
Finally, using  the expansion (Eq. (4) in the paper) that was just  derived for the CSB elemeents, we obtain expansion (5). Equivalently, we can derive (5)  starting from \eqref{Fockn1n2seq}, expanding the product bases $\ket{s}$ in terms of the Schur transform bases, and use the relation \eqref{rels2} together with the reality of the transformation coefficients $\langle j, m , \tau| s \rangle$.

\subsection{General expansion of the Fock basis in terms of the CSB basis}
Although not needed for the results of the paper, it is worthwhile to complete the connection between the Fock basis and the CSB basis. For this we start with
the  the inner product products between the basis elements of the two bases
\begin{equation}
 \langle j, m, m'    |n_{11},n_{12},n_{21},n_{22}\rangle
\end{equation}
and expand $\bra{ j, m, m' }$ using Eq. (...)  and $\ket{n_{11},n_{12},n_{21},n_{22} }$ using
expansion \eqref{Fockgeninseq}, to obtain 
\begin{equation}
\langle j, m, m'    |n_{11},n_{12},n_{21},n_{22}\rangle =
\sqrt{\frac{n_{11}! n_{12}! n_{21}! n_{22}!}{n!\ \gamma_{j,n}} }\sum_{(s, s') \sim (n_{11},n_{12},n_{21},n_{22}) } \langle j,m,\tau|s \rangle
     \langle j,m',\tau|s' \rangle.
\end{equation}
Next we note from the reality of the transformation coefficients $\langle j,m,\tau|s \rangle$ that the sum is precisely the matrix element $\bra{s}X_{j,m,m'}\ket{s'}$ of Eq. \eqref{Xmatel};  hence, from that same equation,
\begin{equation}
\langle j, m, m'    |n_{11},n_{12},n_{21},n_{22}\rangle = 
\sqrt{\frac{\gamma_{j,n}}{\frac{n!}{n_{11}! n_{12}! n_{21}! n_{22}!}}} C_{m,m'}^{j,n}(W)
\end{equation}
where $C_{m,m'}^{j,n}(W)$ is the Louck polynomial as defined by Eq. \eqref{Louckexp}, and $W$ is the integer-values matix with elements $W_{ij}= n_{ij}$. 

\section{Channel probability and its asymptotics}
From Eq. (10) of the paper, the channel probability is given by
\begin{equation}
    p(j|m) = \frac{\gamma_{j,n}}{\binom{n}{\frac{n}{2}+m}} \mathcal{D}^{j,n}_{m,m}(B_q^{\dagger}B_q).
\end{equation}
When expressed in terms of $\eta$,  $p(j|m)$ can be cast in terms of hypergeometric functions or other well-known orthogonal polynomials. Moreover, it exhibits large deviation behavior asymptotically, leading to the concentration-of-measure about the value $j^*$ discussed in the paper. 

\subsection{Derivation of Eq. (12)}
From the definition of the basis change matrix
\begin{equation}
    B_q = \ket{\Gamma_1}\bra{1} + \ket{\Gamma_1}\bra{2},
\end{equation}
we obtain in the orthogonal basis $\ket{1},\ket{2}$ of $\mathcal{H}_q$ the matrix representation
\begin{equation}
    B_q^\dagger B_q \dot{=} \left( \begin{array}{cc} 1 & \langle \Gamma_1 | \Gamma_2 \rangle \\
    \langle \Gamma_2 | \Gamma_1 \rangle & 1
    \end{array}
    \right), 
\end{equation}
and note that
\begin{equation}
\det(B_q^\dagger B_q) = 1 - \eta^2 .
\end{equation}
Defining $\chi= \arg \langle \Gamma_2 | \Gamma_1 \rangle$ and, recalling the definition of $\eta = |\langle \Gamma_2 | \Gamma_1 \rangle|)$, 
we can factorize the matrix as
\begin{equation}
    B_q^\dagger B_q \dot{=} \left( \begin{array}{cc} e^{- i \chi/2} & 0 \\
    0 & e^{ i \chi/2}
    \end{array}
    \right)  \left( \begin{array}{cc} 1 & \eta \\
    \eta & 1
    \end{array}
    \right) \left( \begin{array}{cc} e^{ i \chi/2} & 0 \\
    0 & e^{ -i \chi/2}
    \end{array}
    \right) = e^{-i \chi \sigma_z/2 }( \mathds{1} + \eta \sigma_x) e^{i \chi \sigma_z/2 }
\end{equation}
Since $\mathcal{D}^{j,n}_{m,m}(g)$ is a representation,
\begin{equation}
    \mathcal{D}^{j,n}_{m,m}(B_q^\dagger B_q) 
    = \left[ 
    \mathcal{D}^{j,n}(e^{-i \chi \sigma_z/2 })  \mathcal{D}^{j,n}(\mathds{1} + \eta \sigma_x)\mathcal{D}^{j,n}(e^{i \chi \sigma_z/2 })
    \right]_{m m} \, .
\end{equation}
Moreover, recalling that for $GL(2,\mathbb{C})$ matrices
\begin{equation}
    \mathcal{D}^{j,n}_{m,m}(g) = \det(g)^{\frac{n}{2}-j} {D}^{j}_{m,m}(g),
\end{equation}
where ${D}^{j}_{m,m}(g)$ are the  $SU(2)$ representation matrix elements, we have
$
     \mathcal{D}_{m m'}^{j,n}(e^{\pm i \chi \sigma_z/2 }) = e^{\pm i m \chi } \delta_{m m'}
$ 
and therefore
\begin{equation}
    \mathcal{D}^{j,n}_{m,m}(B_q^\dagger B_q) 
    =  \mathcal{D}_{m m}^{j,n}(\mathds{1} + \eta \sigma_x).
\end{equation}
Next using expansion \eqref{SU(2)polyexp} with $g = \mathds{1} + \eta \sigma_x$ (i.e., $g_{11}=g_{22} = 1$ and $g_{12} = g_{21} =\eta$), we obtain
\begin{eqnarray}
\mathcal{D}^{j,n}_{m,m}(B_q^\dagger B_q) & = & (1-\eta^2)^{\frac{n}{2}-j}
   \sum_{x=0}^{j-|m|} \frac{(j+m)!(j-m)! \eta^{2x}}{(x!)^2 (j+m-x)! (j-m-x)!} \\
   & = & (1-\eta^2)^{\frac{n}{2}-j}
   \sum_{x=0}^{j-|m|} \binom{j+m}{x} \binom{j-m}{x}  \eta^{2x} \label{Dbinform}
\end{eqnarray}
The sum on the right-hand side can be written in terms of Gauss hypergeometric functions as
\begin{equation}
   {}_2F_1(-j-m,-j+m;1;\eta^2)=  \sum_{x=0}^{j-|m|} \binom{j+m}{x} \binom{j-m}{x}  \eta^{2x}  ,
\end{equation}
yielding finally equation (12) in the paper.  With appropriate parameter transformations, the sum can also be expressed in terms of Meixner polynomials or Kravtchuk polynomials, which are also given in terms of ${}_2F_1$ hypergeometric functions (see e.g. \cite{koekoek_askey-scheme_1996}). 

Putting all the terms together, the final   exact expression for the channel probability in terms of Gauss hypergeometric functions reads
\begin{equation}
    p(j|m) = \frac{2j+1}{\frac{n}{2}+j+1}\binom{n}{\frac{n}{2}+j}  {\binom{n}{\frac{n}{2}+m}}^{\! -1} (1-\eta^2)^{\frac{n}{2}-j}\ {}_2F_1(-j-m,-j+m;1;\eta^2).
    \label{fullchanprob}
\end{equation}

\subsection{Derivation of Eq. (20)}
We now turn to the  asymptotics  of $p(j|m)$ and the determination of the critical value $j^*$, which is interpreted asymptotically as half  the number of {effectively} indistinguishable photons interacting at the BS. In the paper, we showed that the OID  admits the expansion 
\begin{equation}
    p(m'|m) = \sum_j p(j|m)p(m'|j,m),
\end{equation}
where $p(j|m)p(m'|j,m)$ comes from the product rule $p(j,m'|m) = p(j|m)p(m'|j,m)$ and $p(j|m)$ is the channel OID for $2 j$ indistinguishable photons.  We now show that $p(j|m)$ exhibits asymptotic concentration  precisely around the value $j^* = n \jb^*$, where  \begin{equation}
    \jb^* =  \sqrt{  \frac{\eta^2}{4} + (1- \eta^2) \mb^2 },
\end{equation}
and $\mb = m/n$, so that in terms of the scaled variables $\mb$ and $\jb = j/n$, the sequence of distributions  
\begin{equation}
    p_n(\overline{j}|\overline{m}) = \sum_{j=0}^{n/2} p(j|\lceil n \mb \rceil )\delta(\jb - j/n)
\end{equation}
satisfies 
\begin{equation}
\lim_{n \rightarrow \infty} p_n(\jb|\overline{m}) = \delta(\jb - \jb^* ). 
\end{equation}
This concentration is a consequence of the fact that  $p(\jb|\mb)$ satisfies a large deviation principle \cite{sornette_critical_2000,touchette_basic_2012}
\begin{equation}
    p(\jb|\mb) \sim e^{-n r(\jb|\mb)},
\end{equation}
with a convex (in $\jb$) rate (or Cramér) function, 
\begin{equation}
   r(\jb|,\mb) = -\lim_{n\rightarrow \infty} \frac{1}{n}\ln p(\, j=n\overline{j}\, |\, m=n\overline{m}\, ),
\end{equation}
for which $\jb^*$ is its minimum. 

To establish the rate function, we turn to the  expression for $p(j|m)$ obtained by using \eqref{Dbinform}, namely
\begin{equation}
  p(j|m) =  \frac{2j+1}{\frac{n}{2}+j+1}\binom{n}{\frac{n}{2}+j}  {\binom{n}{\frac{n}{2}+m}}^{\! -1} (1-\eta^2)^{\frac{n}{2}-j}\sum_{x=0}^{j-|m|} \binom{j+m}{x} \binom{j-m}{x}  \eta^{2x} \, .
\end{equation}
Letting $B(k|N,p)$ be the binomial distribution with parameters $(n, p)$ for a discrete variable $k$ 
\begin{equation}
    B(k|n,p)= \binom{n}{k}p^k (1-p)^{n-k} \, \qquad k, \in \{0, \ldots 1 \}
\end{equation}
we can rewrite the factor $(1-\eta)^{\frac{n}{2}-j}$ as
\begin{equation}
    (1-\eta)^{\frac{n}{2}-j}   = \Big(\frac{1-\eta}{2}\Big)^{\frac{n}{2}-j} \Big(\frac{1+\eta}{2}\Big)^{\frac{n}{2}+j} 2^{\frac{n}{2}+m} 2^{\frac{n}{2}-m} \frac{1}{(1+\eta)^{j+m}(1+\eta)^{j-m}},
\end{equation}
to obtain an expression for $ p(j|m) $ in terms of binomial distributions:
\begin{equation}
  p(j|m) = \kappa \times  B\left(\left. \frac{n}{2} +j\right|n,\frac{1 + \eta}{2}\right) B\left(\left. \frac{n}{2} +m \right|n,\frac{1 }{2}\right)^{\!-1} \sum_{x=0}^{j-|m|}
  B\left( x\left|j+m,\frac{\eta}{1 + \eta}\right. \right)
  B\left( x\left|j-m,\frac{\eta}{1 + \eta}\right. \right)
\end{equation}
where $\kappa =\frac{2j+1}{\frac{n}{2}+j+1}$ is  $O(1)$ asymptotically. The convenience of this expression is that  the binomial distribution has 
a well-known large deviation behavior; namely  \cite{sornette_critical_2000} 
\begin{equation}
    B(k| n, p) \sim e^{- n  \drel{\overline{k}}{p} }
\end{equation}
where the rate exponent $\drel{\overline{k}}{p} = \lim_{n\rightarrow \infty}\frac{1}{n}\ln B(n \overline{k}| n, p)$ is the binary relative entropy
\begin{equation}
    \drel{\overline{k}}{p} := \overline{k} \ln \frac{\overline{k}}{p} + (1-\overline{k}) \ln \frac{1-\overline{k}}{1-p}.
\end{equation}
Therefore the rate function can be written as 
\begin{equation}
    r(\jb|,\mb) = \drel[\bigg]{\frac{1}{2} + \jb}{\frac{1+\eta}{2}}  
    -\drel[\bigg]{\frac{1}{2} + \mb}{\frac{1}{2}} 
    -  \lim_{n \rightarrow \infty} \frac{1}{n} \ln\left[ \ \sum_{x=0}^{\lfloor n(\jb-|\mb|)\rfloor} T(x,j) \right]
\label{ratefunsum}
\end{equation}
where
\begin{equation}
  T(x,j)  =
  B\left( x\left|j+m,\frac{\eta}{1 + \eta}\right. \right)
  B\left( x\left|j-m,\frac{\eta}{1 + \eta}\right. \right).
\end{equation}
The last term of Eq. \eqref{ratefunsum} can be computed using Laplace 's principle: In the sum, all  terms  are  positive and the number of terms in the sum is $j -|m|+1 $; therefore the sum has the two-sided bound
\begin{equation}
  T_{\max}(j)  \leq   \sum_{x=0}^{j -|m|} T(x,j) \leq (j- |m| + 1) T_{\max}(j) \, , \qquad T_{\max}(j) = \max_{0 \leq x \leq j- |m|  }T(x,j) .
\end{equation}
Applying the logarithm on both sides of this inequality, dividing by $n$, and noting that $n (j -|m|+1) $ is $O(\ln(n)/n)$, 
 we get
\begin{equation}
     -\lim_{n\rightarrow\infty} \frac{1}{n} \ln\left[ \ \sum_{x=0}^{\lfloor n(\jb-|\mb|)\rfloor} T(x,j) \right] = -\lim_{n\rightarrow\infty} \frac{\ln(T_{\max}(j))}{n} = -\inf_{\overline{x} \in [0,\jb + \mb]  } \lim_{n\rightarrow\infty} 
     \frac{1}{n}\ln T(n \overline{x},n \jb ) 
\end{equation}
which is the minimum value of the rate exponent of $T(x,j)$. Since the binomial distributions in $T(x,j)$ are parameterized by $j \pm m$ and not $n$, the rate exponent needs to be recast in terms of the growth rate $n$. This gives
\begin{equation}
      -\lim_{n\rightarrow\infty} 
     \frac{1}{n}\ln T(n \overline{x},n \jb )  = (\jb+\mb) \drel[\bigg]{\frac{\overline{x}}{\jb +\mb}}{\frac{\eta}{1 + \eta}}
     + (\jb - \mb) \drel[\bigg]{\frac{\overline{x}}{\jb - \mb}}{\frac{\eta}{1 + \eta}}\, .
\end{equation}
Minimizing the right hand side with respect to $\overline{x}$   gives the equation for the location of the minimum:
\begin{equation}
    \ln \left( \frac{\overline{x}^2 }{(\jb - \mb - \overline{x} )(\jb + \mb - \overline{x} ) \eta^2  }  \right) = 0.
\end{equation}
 Denoting the  minimizer by $\overline{x}_{\jb}$,  we readily find that
\begin{equation}
    \overline{x}_{\jb} = \frac{\eta}{1-\eta^2} \left( \sqrt{\jb^2-(1-\eta^2)\mb^2} - \eta\,  \jb \right).
    \label{eqx}
\end{equation}
Our final expression for the rate function of the channel probability is then:
\begin{equation}
   r(\jb|\mb) =  \drel[\bigg]{\frac{1}{2} + \jb}{\frac{1+\eta}{2}}  
    -\drel[\bigg]{\frac{1}{2} + \mb}{\frac{1}{2}} + 
    (\jb+\mb) \drel[\bigg]{\frac{\overline{x}_{\jb}}{\jb +\mb}}{\frac{\eta}{1 + \eta}}
     + (\jb - \mb) \drel[\bigg]{\frac{\overline{x}_{\jb}}{\jb - \mb}}{\frac{\eta}{1 + \eta}} .
\end{equation}
where $\overline{x}_{\jb}$ is given by Eq. \eqref{eqx}.

To study the concentration of measure of $p(j|m)$ we  look at the  first two derivatives of  $r(\jb|\mb)$ with respect to $\jb$. The first  derivative can be expressed in terms of $\overline{x}_{\jb}$ as
\begin{equation}
    r'(\jb|\mb) = \ln \left[ \frac{1-\eta^2}{ \eta^2} \, \frac{1 + 2 \jb}{1- 2 \jb}\, \frac{x_{\jb}^2}{(\jb^2 -\mb^2)} \right]
    \label{rd1}
\end{equation}
while the second derivative is more transparently expressed as
\begin{equation}
    r''(\jb|\mb) = \frac{4}{1-4 \jb^2} + \frac{ 2 \eta\,  \mb^2 }{(\jb^2 -\mb^2)\sqrt{\jb^2-(1-\eta^2)\mb^2}  }.
    \label{rd2}
\end{equation}
From the second derivative we  verify that $ r''(\jb|\mb)>0 $ in the interval $[|\mb|, 1/2 ]$, and therefore that $r(\jb|\mb)$ is convex. Inserting $\overline{x}_{\jb}$ into Eq. \eqref{rd1}, and solving for $r'(\jb|\mb)=0$ finally gives one sole root in that interval, which is now guaranteed to be a minimum. The root is precisely
\begin{equation}
    \jb^* =  \sqrt{  \frac{\eta^2}{4} + (1- \eta^2) \mb^2 },
\end{equation}
as advertised.

For completeness, it is worthwhile to derive the Gaussian approximation to $p(\jb|\mb)$ about $j^*$. Expanding the rate function about $\jb^*$ to second order:
\begin{equation}
r(\jb|\mb) = r(\jb^*|\mb) + \frac{r''(\jb^*|\mb)}{2}(\jb - \jb^*)^2 + O((\jb - \jb^*)^3),
\end{equation}
one can verify that $r(\jb^*|\mb)=0$ and that
\begin{equation}
r''(\jb^*|\mb) = \frac{16 (\jb^*)^2}{ \eta^2 (1 - \eta^2)(1 - 4 \mb^2)}. 
\end{equation} 
Hence, from $p(\jb|\mb) \sim e^{- n r(\jb| \mb) }$,
we find that asympotically $p(\jb|\mb)$ is Gaussian with
\begin{equation}
\Delta \jb =  \ \frac{\eta \sqrt{1 - \eta^2} \sqrt{1 - 4 m^2} }{4\, \sqrt{n}\, \jb}.
\end{equation}



\section{The OID for perfectly distinguishable photons and the binary summetric channel}
When the photons are perfectly distinguishable ($\eta=0$), the output statistics is consistent with a process in which the photons pass independently through the BS (as for instance when the photons are fed one at a time with no overlap in their temporal wave functions). In this case the statistics can be understood as sampling from a binary symmetric channel (BSC). The correspondence is most transparently established by working with the product basis expansion of $\mathcal{H}_1^{\otimes n} \cong \mathcal{H}_P \otimes \mathcal{H}_Q $ of the Fock states.

As discussed in the paper, when $\eta=0$, the basis transformation matrix $B_q$ is the identity and the initial state is precisely the Fock state $\ket{n_1,0,0,n_2}$, with product basis expansion \eqref{Fockn1n2seq}. The output state is therefore
\begin{equation}
    \ket{\psi}_{out} = \frac{1}{\sqrt{ \binom{n}{n_1}}} \sum_{ s \sim (n_1,n_2)} U_p^{\otimes n}\ket{s}_P \ket{s}_Q.
\end{equation}
Since we are only interested in port counts, we can work with the partial density matrix of output the  ports, 
\begin{equation}
\rho_{P,out} = \tr_Q(\ket{\psi}\bra{\psi}_{out}).
\end{equation}
Since the product basis states are orthonormal, this partial state becomes a mixture single particle product states:
\begin{equation}
\rho_{P,out} = \frac{1}{ \binom{n}{n_1}}\sum_{ s \sim (n_1,n_2)} U_p^{\otimes n}\ket{s}\bra{s} (U_p^\dagger)^{\otimes n}.
\end{equation}
The probability of obtaining $(n_1',n_2'=n-n_1')$ counts at the detectors is then given by adding all diagonal elelements in the product basis of type $(n_1',n_2'=n-n_1)$:
\begin{equation}
p(n_1'|n_1,n_2) = \sum_{s'\sim (n_1',n_2')} \bra{s'}\rho_{P,out}\ket{s} = \frac{1}{ \binom{n}{n_1}}\sum_{s'\sim (n_1',n_2')}\sum_{ s \sim (n_1,n_2)}|\bra{s'} U_p^{\otimes n}\ket{s}|^2,
\end{equation}
This sum is similar to the one encountered in Eq. \eqref{Dsuminseqs}, except that it involves $|\bra{s'} U_p^{\otimes n}\ket{s}|^2$ as opposed to $\bra{s'} U_p^{\otimes n}\ket{s}$; the monomials to collect are therefore in the elements
\begin{equation}
Q_{ij} = |g_{ij}|^2 
\end{equation}
which are precisely the elements of the transition probability $p(i|j)$ for a single photon that enters port $j$ to exit at port $i$. Collecting monomials in the $Q_{ij}$, we obtain
 \begin{equation}
    p(n_1'|n_1,n_2) =     n_1 ! n_2 !  \sum_{W}^*\prod_{ij} \frac{ Q_{ij}^{W_{ij} }   }{ W_{ij}! },
 \end{equation}
where the rows of $W$ add to the $n_i'$ and its columns to the $n_j$. Incorporating these constraints, the sum can finally be written as
\begin{eqnarray}
p(n_1'|n_1,n_2) & = & \sum_{k} \left[\binom{n_1}{k}Q_{11}^{k} Q_{21}^{n_1-k}  \right] \left[\binom{n_2}{n_1'-k}Q_{12}^{n_1'-k} Q_{22}^{n_2-n_1'+ k}  \right] \\ & = & \sum_{k} B(k|n_1, Q_{11}) B(n_1'- k| n_2, Q_{12}),
\label{BSCconvo}
\end{eqnarray}
where $B(k|n,p)$ stands for the binomial distribution as before and we use the fact that since $U_p$ unitary, the row and column sums of $Q_{ij}$ are equal to $1$. Therefore we obtain the discrete convolution of two binomials, which is consistent with interpreting the counts $n_1'$ as the sum of two independent variables $k_1$ and $k_2$:
\begin{equation}
n_1'= k_1 + k_2, \qquad k_1 \sim Bin(n_1, 1-f), \ \ k_2 \sim Bin(n_2, f)
\end{equation}
where 
\begin{equation}
f = Q_{12} = Q_{21}= |\bra{1}U_p\ket{2}|^2 \, .
\end{equation}   
This is precisely the distribution obtained from sampling the BSC $n_1$ times through input $1$ and $n_2$ times from input $2$, where $f$ is the BSC error (or flip) probability. The resulting distribution therefore has mean and variance
\begin{eqnarray}
    \langle n_1'\rangle & = & n_1 (1-f) + n_2 f = \frac{n}{2}  \\
    \langle (\Delta n_1')^2 \rangle & = & n f (1-f) 
\end{eqnarray}
or in terms of imbalances, and  using $f= \sin\theta/2^2$ as in the paper, gives
\begin{eqnarray}
    \langle m'\rangle & = &  m \cos \theta \\
    \langle (\Delta  m)^2 \rangle & = &  \frac{n}{4} \sin^2\theta.
\end{eqnarray}
Finally, from Eq. \eqref{BSCconvo} and  the Laplace principle, it follows that for $\eta=0$, the OID satisfies a large deviation principle 
\begin{equation}
    p(m'|m)_{\eta=0} \sim e^{-n r(\mb'|\mb) }
\end{equation}
with rate function
\begin{equation}
   r(\mb'|\mb) = \inf_{\overline{k} \in[0,\frac{1}{2} + \mb]}
    \left[\left(\frac{1}{2}+ \mb\right)\drel[\bigg]{ \frac{\overline{k}}{\frac{1}{2}+ \mb} }{1-f} + \left(\frac{1}{2}- \mb\right)\drel[\bigg]{ \frac{\frac{1}{2}+\mb'-\overline{k}}{\frac{1}{2}-\mb} }{f}\right]\, .  
\end{equation}

\bibliographystyle{IEEEtran}
\bibliography{supplemental}